\documentclass[runningheads]{llncs}
\usepackage{graphicx}
\usepackage{hyperref}
\usepackage{listings}
\begin{document}

    \title{Dataset Lifecycle Framework and its applications in Bioinformatics}

    \titlerunning{DLF and its applications in Bioinformatics}
    \author{Yiannis Gkoufas\inst{1} \and
    David Yu Yuan, Ph.D.\inst{2}\orcidID{0000-0003-1075-1628}}

    \authorrunning{Y. Gkoufas et al.}
    \institute{IBM Research - Ireland
    \email{yiannisg@ie.ibm.com}
    \url{https://www.ibm.com/} \and
    Technology and Science Integration, European Bioinformatics Institute, European Molecular Biology Laboratory
    \email{davidyuan@ebi.ac.uk}
    \url{https://www.ebi.ac.uk/}}

    \maketitle              % typeset the header of the contribution

    \begin{abstract}
%        \begin{remark}
%            DY + YG: The abstract should briefly summarize the contents of the paper in 150--250 words.
%        \end{remark}

        Bioinformatics pipelines depend on shared POSIX filesystems for its input, output and intermediate data storage.
        Containerization makes it more difficult for the workloads to access the shared file systems.
        In our previous study, we were able to run both ML and non-ML pipelines on Kubeflow successfully.
        However, the storage solutions were complex and less optimal.
        This is because there are no established resource types to represent the concept of data source on Kubernetes.
        More and more applications are running on Kubernetes for batch processing.
        End users are burdened with configuring and optimising the data access, which is what we have experienced before.

        In this article, we are introducing a new concept of Dataset and its corresponding resource as a native Kubernetes object.
        We have leveraged the Dataset Lifecycle Framework which takes care of all the low-level details about data access in Kubernetes pods.
        Its pluggable architecture is designed for the development of caching, scheduling and governance plugins.
        Together, they manage the entire lifecycle of the custom resource Dataset.

        We use Dataset Lifecycle Framework to serve data from object stores to both ML and non-ML pipelines running on Kubeflow.
        With DLF, we make training data fed into ML models directly without being downloaded to the local disks, which makes the input scalable.
        We have enhanced the durability of training metadata by storing it into a dataset, which also simplifies the set up of the Tensorboard, separated from the notebook server.
        For the non-ML pipeline, we have simplified the 1000 Genome Project pipeline with datasets injected into the pipeline dynamically.
        In addition, our preliminary results indicate that the pluggable caching mechanism can improve the performance significantly.

    \end{abstract}

    \section{Introduction}\label{sec:introduction}

    Research pipelines make extensive use of HPC for batch processing.
    They all require direct access to POSIX file systems locally.
    Ever-increasing demands of computational resources by Bioinformatics pipelines make migration to clouds necessary.
    In our previous studies~\cite{ref_lncs1,ref_proc1,ref_article1}, we ported the pipelines from Platform Load Sharing Facility (LSF) to Kubernetes (K8S)~\cite{ref_url1} and Kubeflow (KF)~\cite{ref_url2} to make the pipelines more efficient by running them in containers.
    This made it harder to access cloud-agnostic global shared file systems similar to ones in HPC clusters due to limited choice of persistent volumes~\cite{ref_url3} supporting RWX access mode.
    We made many attempts to provide seemingly local access to file systems required by Bioinformatics tools designed for HPC in our studies\cite{ref_lncs1,ref_proc1,ref_article1}.
    We were never satisfied with the performance, security and ease of use with the previous solutions.

%    \begin{remark}
%        YG: Please update the following.
%    \end{remark}
    Dataset Lifecycle Framework (DLF)~\cite{ref_url4} is a Kubernetes framework which enables users to access remote data sources via a mount-point within their containerized workloads.
    It is extensible and can potentially support additional storage mechanisms, thanks to the bundled Container Storage Interface (CSI)~\cite{ref_url5} plugins.
    It is aimed to improve usability, security and performance with a higher level construct: Dataset, implemented as a Custom Resource Definition (CRD).
    Each Dataset is a pointer to an existing remote data source and is materialized as a Persistent Volume Claim~\cite{ref_url11}
    For Bioinformatics pipelines for both Machine Learning (ML) and classic algorithms (non-ML), we store input data and output results in object stores for nearly unlimited scalability.
    It is common knowledge that object stores do not perform well compared with NAS or local storages.
    Bioinformatics tend to have smaller numbers of larger data files (e.g.\ thousands of files with the size of MiB or more such as genomic sequences, research literature and images).
    DLF can provide a nice POSIX-like facade for Bioinformatics pipelines for sequential reads and writes.

%    \begin{remark}
%        DY + YG: Novelty / results of pipelines with DLF: function, security and performance.
%
%        DLF is not competing with CSI, instead it aims to its adoption as it makes them more easy to use for the non-advanced Kubernetes users.
%    \end{remark}

    We are focusing on the usability, security and performance of DLF when it is used to support Bioinformatics pipelines on Kubeflow.
    We have conducted functional studies on Google Kubernetes Engine (GKE) and performance benchmarking on IBM Cloud Kubernetes Service (IKS).
    Through both ML and non-ML Bioinformatics pipelines, we have investigated the capabilities of DLF to access input from GCS, S3 and web server with S3 and H3 protocols in a secure manner.
    We have managed to simplify our pipelines on Kubeflow via all three usage scenarios: creating dataset statically, injecting mount points with ConfigMap and creating dataset dynamically.
    We have made detailed comparison to demonstrate that the integration of DLF has improved the functionality, security and performance of the pipelines.
    In addition, we have benchmarked our variant calling pipeline for the 1000 Genomes Project accessing large number of genome assemblies via DLF with and without cache.

    With the integration of DLF on Kubeflow, we were able to mount object stores as Persistent Volumes Claims and present them to Bioinformatics pipelines as a POSIX-like file system.
    The developers of Bioinformatics Pipelines can focus on the methodologies and the results of the experiments and not on installing and tuning CSI Plugins, since DLF does that for them.
    In addition, DLF makes use of Kubernetes access control and secret so that pipelines do not need to be run with escalated privilege or to handle secret keys.
    This makes the platform more secure.
    Under the hood, DLF can accelerate data access by transparently caching Datasets which are frequently accessed.
    We have achieved all three objectives: usability, security and performance by using DLF to access data for Bioinformatics pipelines on Kubeflow.
    The Dataset Lifecycle Framework is designed to be cloud-agnostic, depending on a Kubernetes cluster only.
    The conclusion drawn from the study is applicable to any cloud with an preexisting Kubernetes cluster although the detailed benchmarks would likely vary due to environmental factors such as the amount of resources available to the cluster in the clouds.

%    \begin{remark}
%        DY + YG: Three cloud environments: IKS, EKS and GKE. At least 2: IKS and GCP\@.
%
%        Functionally working on three platforms.
%        Performance benchmarking on IKS only.
%        YG: Definitely agree with that, its the only
%        Avoid the extra work and potential controversy for benchmarking on all three clouds.
%    \end{remark}

    \section{Method}\label{sec:method}
% 
%    \begin{remark}
%        DY + YG: Some potential discussion points: From s3fs to goofys.
%        YG: It's difficult to have any comparison with s3fs since it hangs and tries to download the entire genome dataset.
%            We can try with their recent version but I doubt that they fixed it.
%        Caching: Rook / Ceph vs goofys
%        YG: Some news on that, I actually managed to have caching with https://www.noobaa.io/ which is more what IBM wants to push for.
%        And its part of their functionality, while the Ceph cache is just our fork.
%        Basically we can compare 1)goofys simple 2)goofys with caching flags 3)noobaa caching
%        DY: I agree Yiannis.
%        Let's forget about s3fs.
%        We focus on goofys and noobaa as you said.
%    \end{remark}
%
    Dataset Lifecycle Framework aims at providing a higher level of abstraction for dynamic provisioning of storage for the users' applications.
    It introduces the concept of Dataset to represent available storage or pre-existing data provided by any cloud-based storage solution, like S3 Object Store or NFS\@.
    In the following sections we will outline the basic specifications and components of the framework.

    \subsection{Dataset Custom Resource Definition}\label{subsec:dataset-crd}
\begin{figure}
        \begin{center}
            \includegraphics[scale=0.5]{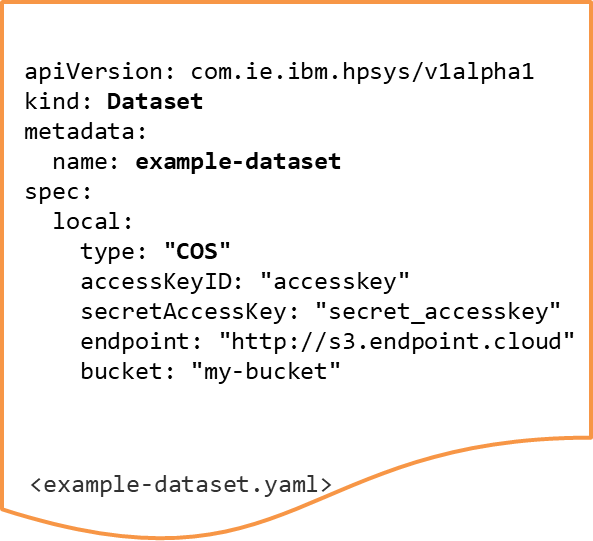}
            \caption{Dataset Spec}\label{fig:fig7}
        \end{center}
    \end{figure}

    Every entity (pod, deployment, job etc) in the Kubernetes world can be expressed as a resource accompanied with a specification.
    As an example, when the user wants to create a new pod, they need to create a pod resource following the related specification in terms of mandatory fields, possible values etc.
    The default Kubernetes resources are materialized by Kubernetes in a well-defined way.
    For instance, when the user creates a new pod, the associated containers will be launched in one of the available nodes of the cluster.
    The mechanisms of Kubernetes enable service providers to define their own Custom Resource Definitions (CRD) to bring the domain-specific logic they desire to their Kubernetes-managed infrastructure.
    Tapping on those capabilities, the Dataset Lifecycle Framework introduces the Dataset CRD\@.
    A Dataset object is a reference to a storage provided by a cloud-based storage solution, potentially populated with pre-existing data.
    Our framework is completely agnostic to where/how a specific Dataset is stored, as long as the endpoint is accessible by the nodes within the Kubernetes cluster, in which the framework is deployed.
    Currently, the framework supports any Cloud Object Storage solution which exposes an S3 or NFS compatible endpoint.
    An example Dataset is displayed in Fig.~\ref{fig:fig7}.

    The user can specify the type of the remote data storage.
    For Datasets stored in COS (Cloud Object Store) the mandatory fields are: \verb|endpoint|, \verb|bucket|, \verb|accessKeyID| and \verb|secretAccessKey|.
    As described above, the only requirement for the endpoint is that it must be reachable address from the Kubernetes nodes.
    The bucket entry creates a one-to-one mapping relationship between a Dataset object and a bucket in the Cloud Object Store.
    The fields \verb|accessKeyID| and \verb|secretAccessKey| are the necessary credentials to allow access to the specific bucket.
    Finally, in the specific example the user or an administrator who creates this Dataset object defines a name for the Dataset, in this case \verb|example-dataset|.
    In Kubernetes the various resources are logically grouped under a namespace, and the users who have access granted to a specific namespace can access all the resources created within it.
    This is valid for the Dataset objects as well, access is restricted to only a specific namespace.
    To accelerate the development process of Kubernetes-based applications and workloads, it is very likely that a different persona would create a Dataset object (like the administrator or the data provider) and a different persona would use a Dataset within their application.

    \subsection{Dataset Operator}\label{subsec:dataset-operator}

    We have described above the specification of a Dataset object.
    Creating a Custom Resource Definition is just the first step to add custom logic in the Kubernetes cluster.
    The next step would be to create a component that has embedded the domain-specific application logic for the CRD\@.
    Essentially, a service provider needs to develop and install a component which reacts to the various events which are part of the lifecycle of a CRD (creation, deletion) and implements the desired functionality.
    This would have required significant amount of boilerplate code and there was no established standard for developing such component a few years ago.
    Fortunately, the opensource operator-sdk [TODO add link] provides the necessary tooling and automation to assist in the development of these components.
    We have utilized the operator-sdk to create our Dataset Operator.
    Its main functionality is to react to the creation of a new Dataset and materialize the specific object.

\begin{figure}
        \includegraphics[width=\textwidth]{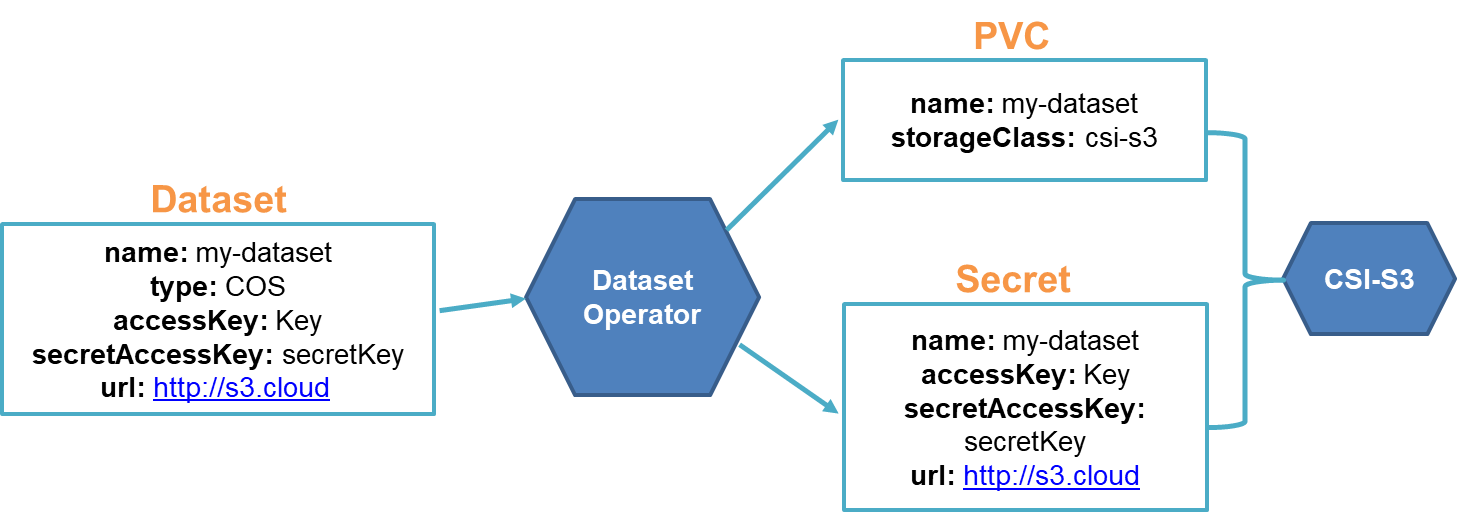}
        \caption{Dataset Operator}\label{fig:fig8}
    \end{figure}

    As mentioned above, the way to provide access to a data source within an application running on Kubernetes is via PVCs (Persistent Volume Claim).
    A PVC is linked to a specific \verb|storage class| corresponding to the underlying storage mechanism.
    However, the support for storage is not flexible when it comes to provisioning new volume types.
    Recently Kubernetes has adopted the Container Storage Interface (CSI)~\cite{ref_url5} that facilitates the introduction of volume plugins without the need of modifying the core of Kubernetes itself.
    For our work we are maintaining an opensource implementation of CSI-based volumes for S3.
    When a new dataset of type COS (Cloud Object Store) is created, the operator is the component responsible for 1) creating a new PVC with the corresponding Storage Class and 2) creating the Kubernetes secrets to store the access credentials for the new Dataset.
    The pair PVC-Secret associated with this Dataset is passed to the CSI-S3 implementation which provides a mount point as a way for any pod to consume the Dataset.
    An example flow is presented in Fig.~\ref{fig:fig8}, where hexagons represent components of the framework and the squares are the Kubernetes resources created as a result of submitting a new Dataset.

    \subsection{Pods Admission Controller}\label{subsec:pods-admission-controller}

\begin{figure}
        \includegraphics[width=\textwidth]{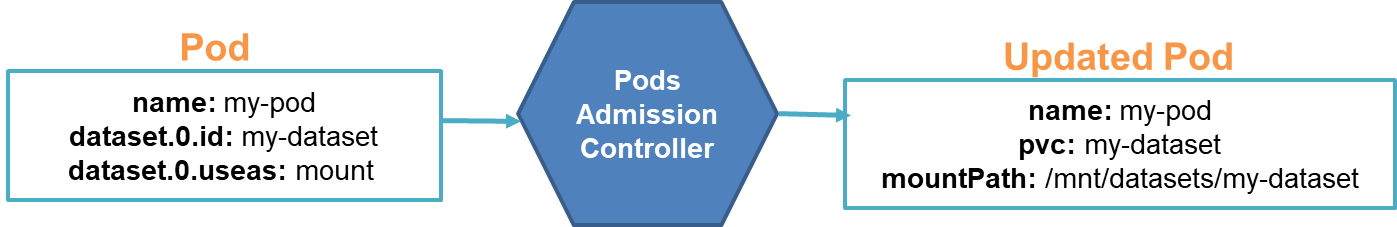}
        \caption{Pods Admission Controller functions diagram}\label{fig:fig9}
    \end{figure}

    With the components we have outlined so far, we enable the end user to define a Dataset representing a remote data source and we create the associated PVC\@.
    This PVC can be mounted to any pod and provide access to a deployed application.
    We have taken an extra-step to offer a transparent way for the users to access their Datasets within their pods.
    With our framework the only requirement from the user is to annotate their pods according to our convention and the Dataset will become available to the pod via a mount point as shown in Fig.~\ref{fig:fig9}.

\begin{figure}
        \begin{center}
            \includegraphics[scale=0.5]{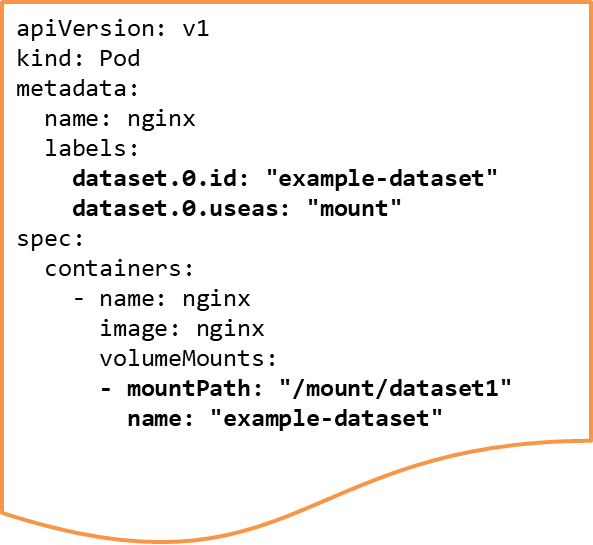}
            \caption{Pod definition with user defined dataset mountpoint}\label{fig:fig10}
        \end{center}
    \end{figure}

    The user needs to define the fields \verb|id| and \verb|useas| for every dataset.
    The \verb|id| field should correspond to a dataset name (as created above).
    The \verb|useas| dictates the way that the user wants to access the specific dataset.
    The user can access it via a mount directory within their pod, but also it's possible for users to get the datasets access credentials (e.g., S3 ID and KEY) injected into application pods.
    Access to data in this latter case will be explicitly handled by the application code using whatever native interface exposed by the dataset source (e.g., S3).
    The extra labels required to operate with datasets are handled by a dedicated Kubernetes component following the Admission Controller pattern.
    Such components can modify objects upon creation and decorate them with additional fields.
    In our use case, we monitor the pods with the specific type of labels and add the necessary information to enable pods mounting the PVC linked to each Dataset.
    As a convention, a dataset with id \verb|my-dataset| would be mounted inside the pod on \verb|/mnt/datasets/my-dataset|.
    The user is also able to override the desired mount path as shown in Fig.~\ref{fig:fig10}

    \section{Result}\label{sec:result}
    We have been using Kubeflow as an efficient platform for both ML and non-ML pipelines in Bioinformatics.
    The pipelines are unable to read from and write to object stores or to access HTTP archives directly.
    They all assume the local access of files on shared file systems.
    It is often impractical to download or upload large amount of files or to rewrite the pipelines to make use of the cloud-native storages.
    We use DLF to bridge the gap.

    \subsection{Server-side}\label{subsec:server-side}

%    \begin{remark}
%        DY + YG: Potential points: authentication, access control.
%        s3, nfs, ceph?
%        archive?
%        YG: What do we want to describe here?
%        The various types of datasets?
%        DY: Yes, I think that we need to give an overview of DLF.
%        Its scope, capability, pros and cons.
%        We save the future plans in discussion section near the end.
%    \end{remark}

%    \begin{remark}
%        DY: examples of public buckets on AWS\@.
%    \end{remark}

%    \begin{remark}
%        DY: EBI public FTP site is very slow.
%        Need to check public HTTP site as well.
%    \end{remark}

\begin{figure}
        \includegraphics[width=\textwidth]{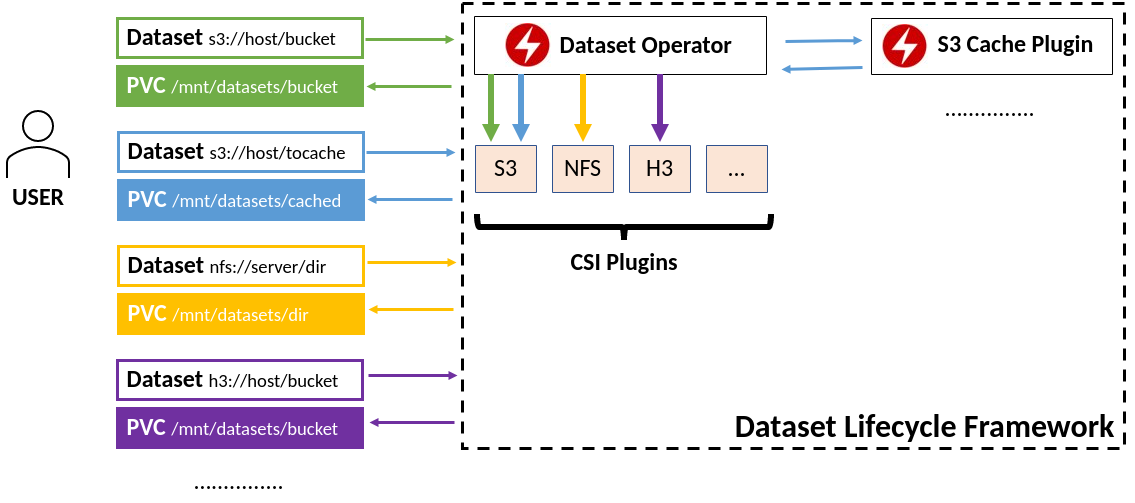}
        \caption{Dataset types and protocols supported by Dataset Lifecycle Framework.}\label{fig:fig11}
    \end{figure}

    DLF supports a number of dataset types: S3, NFS and H3, as shown in Fig.~\ref{fig:fig11}.
    We choose a private Google Cloud Storage (GCS) bucket for cardiomyocyte images for the ML pipeline.
    We use a public S3 bucket on AWS for DNA alignments from the 1000 Genomes Project for the non-ML pipeline.
    We also access a public gzip archive at European Bioinformatics Institute (EBI) for human reference genome GRCh37 for the non-ML pipeline.
    The S3 access is authenticated with the access key and secret. [TODO: access control with Kubernetes service account]
    DLF enables us to access very different input from different sources via different protocols uniformly.
    The data is made available to both ML and non-ML pipelines in containers as if local POSIX files mounted onto pods via PVC dynamically.

    \subsection{Client-side}\label{subsec:client-side}
    Datasets can be mapped into Persistent Volume Claims (PVCs) and ConfigMaps.
    There are three methods to access datasets.
    The three ways of accessing Datasets fit nicely in the three scenarios below:

    \begin{remark}
        YG: please revise the list below.
        I don't think that I have described them well.
%        YG: In this section I think we should focus on how Datasets can be used in the Kubeflow context, what do you think?
%        DY: Totally agree.
    \end{remark}

    \begin{enumerate}
        \item Method 1 - Mount Persistent Volumes via Persistent Volume Claims created with Datasets (see ML: Data Volumes for Notebook Servers in~\ref{subsubsec:notebook}.)%
        \item Method 2 - Mount Datasets via ConfigMaps with pod labels (see ML: Pod labels for Tensorboard in~\ref{subsubsec:tensorboard}.)
        \item Method 3 - Mount Datasets dynamically in a pipeline via Kubeflow Pipeline (KFP) Domain Specific Language (DSL)~\cite{ref_url6} (see Non-ML: PVCs for Kubeflow Pipelines APIs in~\ref{subsubsec:pipeline}.)
    \end{enumerate}

    \subsubsection{ML: Data Volumes for Notebook Servers}\label{subsubsec:notebook}

    Before the integration with DLF, we used Keras API \verb|tf.keras.utils.get_file()| to download files to the local file system~\cite{ref_url7} so that the microscopic images can be used for training and validation later in the notebook.
    It was very slow to download one file at a time via the OMERO 5.6.0 JSON API~\cite{ref_article2}.
    The Keras API \verb|get_file()| provided a simple-minded caching mechanism of skipping files with the same name.
    This can speed up the process but was error-prone.

    With DLF, we created our own repository of images deposited by Nirschl, J.J., et al~\cite{ref_url8} at Image Data Repository (IDR)~\cite{ref_url9}.
    The repository is a storage bucket on GCS via the following process.

%    \begin{remark}
%        YG: Can the above be hosted on IBM Cloud?
%        DY: Yes, we can.
%        If cost is not a concern, we should make a copy of 1000 Genomes bucket from AWS to ICS\@.
%        The full bucket is 700 TB. The objects we need is 230 TB\@.
%        YG: It's better to avoid using 230 TB in IBM Cloud Object store. Maybe we can use it for intermediate data?
%    \end{remark}

\begin{figure}
        \includegraphics[width=\textwidth]{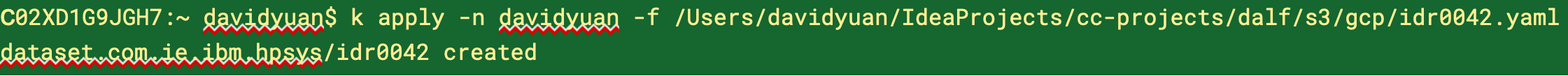}
        \caption{In step 2, A dataset was created to access GCS bucket idr0042 as the private repository for microscopic images cached from IDR.}\label{fig:fig1}
    \end{figure}

\begin{figure}
        \includegraphics[width=\textwidth]{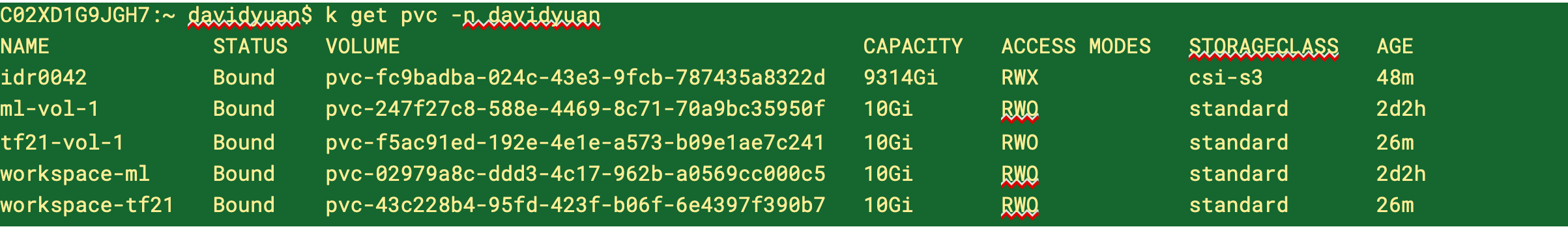}
        \caption{After step 4, a notebook ml was recreated to mount idr0042 in addition to its original ml-vol-1 and workspace-ml. A new notebook tf21 was created to mount idr0042 at the same time.}\label{fig:fig2}
    \end{figure}

    \begin{enumerate}
        \item (GCS) Create an empty single-region bucket, which sacrifices high availability and disaster recovery in favor of high performance and low cost.
        \item (K8S) Define a dataset backed by the newly created bucket in the namespace used by notebooks via kubectl (see Fig.~\ref{fig:fig1}).
        \item (KF) Create a notebook server mounting the dataset via its PVC (method 1).
        \item (Jupyter) Create a Jupyter notebook to download the images onto the data volume  (see Fig.~\ref{fig:fig2}).
    \end{enumerate}

    With an one-time setup, all notebooks on the notebook server can access the images directly from the data volume as if they were on a POSIX filesystem.
    They do not need to download their own copies of images.
    In addition, the GCS bucket can survive the deletion of the notebook, the notebook server, the Kubeflow instance and the Kubernetes cluster.
    The process described above was only needed once.
    The private repository can be used by new notebooks on new Kubeflow instances or by multiple Kubernetes clusters at the same time.

    DLF can not only simplify the implementation of ML pipelines but also reduce the time by eliminating the repeated downloads.
    For the cardiomyocyte image classification notebook, it took almost 11 minutes to download all the 1978 images onto the local /tmp directory.
    If the images were cached (a.k.a.\ Files with the same names exist on the target), it still needed 44 seconds for the overhead of REST calls.
    The notebook accessing the DLF dataset did not make any REST calls.
    The time to retrieve the images was reduced from 11 minutes to 0.

    \begin{table}
        \caption{Dataset Lifecycle Framework eliminating the downloading of microscopic images.}\label{tab:tab1}
        \begin{tabular}{|l|l|l|}
            \hline
            Download without cache (seconds) & Download with cache (seconds) & DLF (seconds) \\
            \hline
            643.0786353652366                & 44.282696074806154            & 0             \\
            \hline
        \end{tabular}
    \end{table}

    It is important to note that the total size of the images downloaded from IDR is 58 MiB, which is quite small.
    ML pipelines typically process GiBs to TiBs of images in training.
    For ML in timelapse videos (e.g.\ embryo development), the data size gets into TiB to PiB range.
    Downloading images via REST calls can quickly become impractical for time and storage space.
    The Datasets backed by object storage provide nearly infinite scalability.
    In addition, ML pipelines prototyped on POSIX file systems can access images in object stores conveniently without modification.

    \subsubsection{ML: Pod labels for Tensorboard}\label{subsubsec:tensorboard}

    Default notebook server images have included Tensorboard as a notebook extension.
    It requires four steps to use it in a notebook in addition to invoke the tensorboard callback:

    \begin{enumerate}
        \item Load the extension in the notebook: \verb|%load_ext tensorboard|
        \item Start the tensorboard in the notebook: \verb|%tensorboard --logdir='<log_dir>'|
        \item Enable port forward in a command terminal on the local client:
        \subitem \verb|kubectl port-forward -n <namespace> <notebook_pod> 8080:6006|
        \item Access the tensorboard in a web browser: \verb|http://localhost:8080/|
    \end{enumerate}

    This approach does not involve DLF. We want to start a tensorboard with DLF to solve the following issues with this approach:

    \begin{enumerate}
        \item The process is very manual and inconvenient.
        \item The train metadata can not survive notebook or kubernetes cluster deletion.
        \item This approach would not work if a custom image does not have the Tensorboard extension.
    \end{enumerate}

    We take a fairly new Tensorflow image \verb|tensorflow/tensorflow:2.2.1-py3| and wrap it into a Deployment.
    The tensorboard gets exposed via a Service and an Istio CRD VirtualService so that it can be accessed via a custom URI from the same host as the Kubeflow dashboard.
    More importantly, the Deployment is configured with two labels so that pods can access a Dataset backed by an object store for the tensorboard (method 2).

%    \begin{figure}
%        \includegraphics[width=\textwidth]{figure-3.png}
%        \caption{DLF uses dataset.0.id and dataset.0.uses to mount a PVC.}\label{fig:fig3}
%    \end{figure}

    A Deployment is labelled with \verb|dataset.0.id| and \verb|dataset.0.uses| to get DLF to mount a Dataset with the default mount point \verb|/mnt/datasets/<dataset_name>/| similar to Fig.~\ref{fig:fig10}.
    The training metadata from a notebook \verb|idr0042| for the cardiomyocyte image classification project idr0042 is persisted in an object store \verb|idr0042|.

    In addition, the namespace for the datasets has to be labelled first for DLF to monitor the lifecycle of the datasets in the namespace.

    \begin{lstlisting}[label={lst:lstlisting2}, language=bash]
    kubectl label namespace "${namespace}" \
    monitor-pods-datasets=enabled --overwrite=true
    \end{lstlisting}

    We use Kustomize to make sure that a tensorboard is created specifically for a notebook using a specific dataset in a particular namespace in our Bash script.
    We are following a convention to use a separate dataset for each notebook server for long term tracking purposes.
    With DLF, Kustomize and a Bash wrapper, we can create separate tensorboards for our microscopic image classification notebooks easily.
    The metadata generated during the training can be preserved in object stores.
    All three issues are resolved.
    Again, DLF simplifies the use of Kubeflow.

%    \begin{remark}
%        DY: picture of Tensorboard
%    \end{remark}

    \subsubsection{Non-ML: PVCs for Kubeflow Pipelines APIs}\label{subsubsec:pipeline}

    Kubeflow Pipeline can access existing Datasets via PipelineVolumes.
    This is similar to how Persistent Volumes are accessed statically in Kubernetes.

    \begin{enumerate}
        \item Create a Dataset manifest (e.g.\ \verb|<my_dataset>.yaml|).
        \item Create a Dataset object via \verb|kubectl apply -f <my_dataset>.yaml| for example.
        \item Access the dataset as \verb|pvolume = PipelineVolume(pvc=<dataset_name>)| in KFP Python script.
    \end{enumerate}

    A more convenient approach is to define Datasets dynamically, again similar to how PVCs can be used to create PVs dynamically (method 3).
    Here are the steps:

    \begin{enumerate}
        \item Create a Dataset manifest (e.g.\ \verb|<my_dataset>.yaml|).
        \item Create a CustomResourceDefinition via \verb|ResourceOp| from the YAML file in the KFP Python script.
        \item Access the dataset in via PipelineVolume.
    \end{enumerate}

    The overall processes for both static and dynamic approaches are similar.
    The dynamic approach gives more control to the pipeline users, similar to how PVCs are used statically or dynamically.
    The code snippet below shows how we have implemented the dynamic approach with the KFP APIs:

    \begin{figure}
        \includegraphics[width=\textwidth]{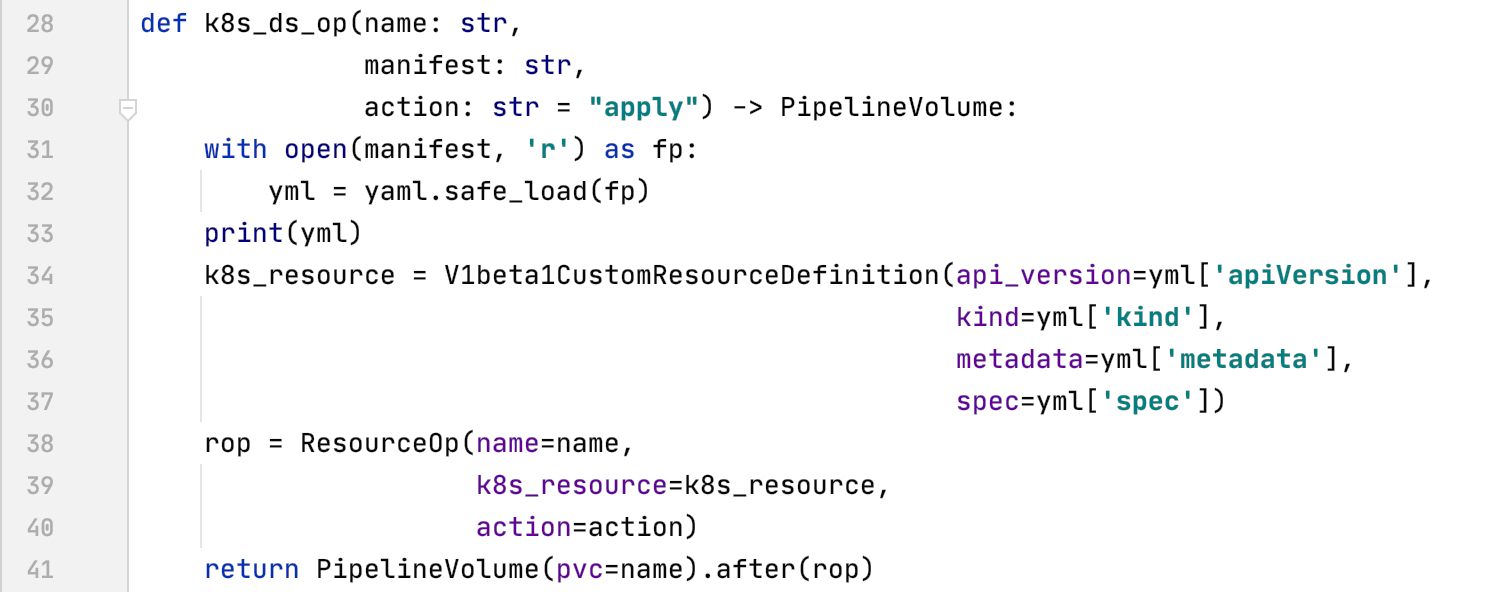}
        \caption{DLF dataset gets created dynamically.}\label{fig:fig4}
    \end{figure}

    With the integration of DLF with KFP APIs in Fig.~\ref{fig:fig4}, we can simplify and improve our existing pipeline for the 1000 Genomes Project.
    The details are discussed below.

    \subsection{End-to-end}\label{subsec:end-to-end}

    \subsubsection{Pipeline for 1000 Genomes Project Simplified}\label{subsubsec:pipeline-for-1000-genome-project-simplified}

    Before the integration with DLF, the output was uploaded to an S3 bucket.
    The input of our G1K pipeline was downloaded from three different sources to PVCs backed by an NFS server for ReadWriteMany (RWX) access:

    \begin{enumerate}
        \item Human reference genome on a public FTP downloaded via curl or wget.
        \item G1K queries, a list of 1046 alignments (i.e.\ BAM file names) on a private S3 bucket downloaded via a custom AWS CLI Docker container.
        \item G1K genomes, the actual BAM files and their indexes (i.e.\ BAI files) on a private NFS volume accessed via Onedata.
    \end{enumerate}

\begin{figure}
        \includegraphics[width=\textwidth]{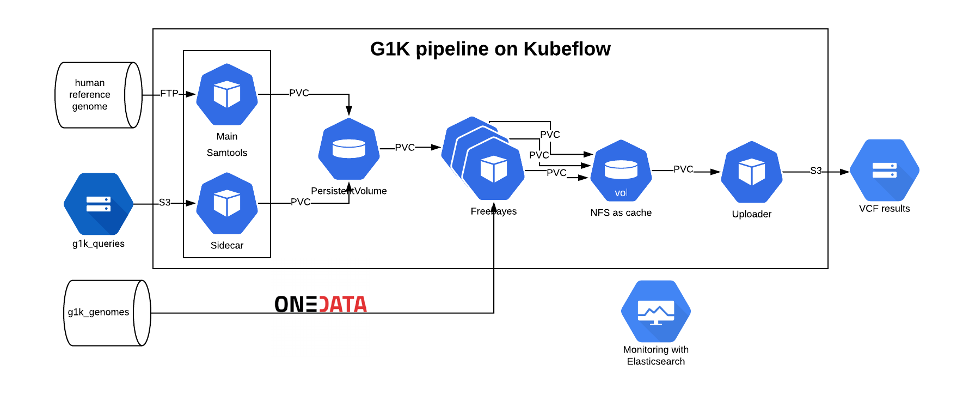}
        \caption{G1K pipeline implementation without DLF with complicated solutions for input, output and temporary storage.}\label{fig:fig5}
    \end{figure}

    The total size of alignments are about 230 TiBs.
    It is impractical to download such a large volume of data and store it on the NFS server.
    It is estimated to take 14 hours to transfer the files and \$2,000 / day to store them.
    We use Onedata~\cite{ref_url10} to access the alignments when they are needed as shown in Fig.~\ref{fig:fig5}.
    As we have discussed in detail in our previous paper~\cite{ref_lncs1}, we have spent very significant effort (compiling Freebayes from the source and merging images via multi-stage builds) to create a custom Docker image because Onedata does not support Kubernetes.
    We also have a maintenance issue to keep updating the Onedata client and Freebays whenever these components change, compounded by poor backward compatibility of Onedata.
    In addition, Onedata client requires root privileges to mount file systems at runtime.
    This implies significant security risk for these custom pods to run with the escalated privilege.

\begin{figure}
        \includegraphics[width=\textwidth]{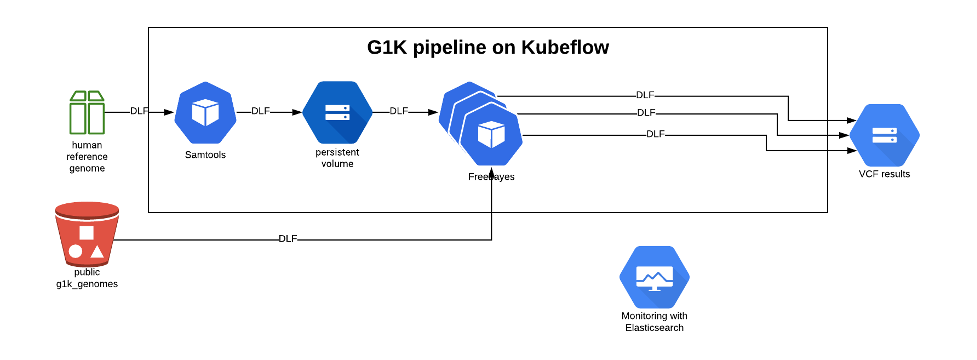}
        \caption{G1K pipeline implementation with DLF with much simplified solutions for input, output and temporary storage.}\label{fig:fig6}
    \end{figure}

    DLF has enabled us to simplify our G1K pipeline significantly (See Fig.~\ref{fig:fig6}).
    We do not download or upload files from or to storage buckets with S3 protocol.
    The sidecar and uploader in Fig.~\ref{fig:fig5} are removed.
    In addition, the process in Freebayes can copy results to the output directly.
    We no longer need an NFS Persistent Volume in RWX mode to cache the results for the uploader.
    These processes can use their local ephemeral disk for caching, which has improved IO throughput very significantly.
    This is because the processes are writing to local disks with completely isolated parallel IOs.
    They no longer have to fight for the shared Persistent Volume backed by an NFS server.
    Finally, Onedata client is no longer used.
    We can simply use the binaries of Freebayes maintained by Anaconda to augment Miniconda image and run pods without escalated privileges.

    Comparing the two implementations of our G1K pipeline, we can definitively conclude that DLF has simplified the implementation significantly.
    In addition, we have improved security by running pods as non-root by removing the dependency on Onedata.
    We can also confidently conclude that the new implementation would perform better by eliminating a bottleneck of the NFS Persistent Volume, and eliminating a custom S3 uploader.
    We will analyse the performance of the new implementation with and without caching for DLF below.

    \subsubsection{Performance enhancements}\label{subsubsec:performance-enhancements}

    \begin{remark}
        DY: Tested upload speed with dd on GCP without caching.
        Read cache vs write cache?
        \begin{lstlisting}[label={lst:lstlisting}, language=bash]
ls -l /tmp/awscli-bundle.zip && dd if="/tmp/awscli-bundle.zip" \
of="${HOME}/ref1000g/dummyfile"  bs=1024 count=16384 \
&& rm "${HOME}/ref1000g/dummyfile"
-rw-r--r-- 1 root root 15528040 Feb  5  2020 /tmp/awscli-bundle.zip
15164+1 records in
15164+1 records out
15528040 bytes (16 MB, 15 MiB) copied, 2.32823 s, 6.7 MB/s
        \end{lstlisting}

    \end{remark}

    \begin{remark}
        DY + YG: GKE, EKS and IKS - DLF caching enabled on all three.
        Performance benchmarked on IKS\@.
        Clarify not comparing performance between clouds.
    \end{remark}

    \section{Discussion}\label{sec:discussion}

    \begin{remark}
        YG: [Some contents from https://github.com/IBM/dataset-lifecycle-framework/wiki/Roadmap is worth mentioning here.]
    \end{remark}

    We have decided not to compare the performance of the two versions of the variant calling pipeline with Onedata and with DLF in this study.
    This is because the architecture of the pipeline has changed significantly.
    Thus, we are unable to compare these two utilities side-by-side to draw meaningful conclusions.
    We have some empirical evidence that the simplification of the pipeline has resulted in the improvements of the overall throughput.

    This paper only discussed the application of Dataset Lifecycle Framework in Bioinformatics.
    However, we believe that DLF would work just as well with other research pipelines by simplifying the integration with cloud-native object stores.
    If pipelines require shared global POSIX file systems especially for sequential reads and writes, DLF can be helpful to ease the transition for such applications from HPC to clouds.

    \section{Conclusion}\label{sec:conclusion}

%    \begin{remark}
%        DY + YG: Simplicity, security and performance.
%    \end{remark}

    Dataset Lifecycle Framework provides a high level construct based on CSI.
    It introduces a concept of dataset to represent available storage or existing data backed by cloud-native storage.
    In particular, it creates Dataset CRDs on Kubernetes to describe the remote data storage.
    It uses Dataset Operator to manage the lifecycles of the CRDs.
    As a result, the corresponding PVCs are created to present the COS to the containers in pods such that application can access the remote data directly.
    In addition, Pods Admission Controller injects the mount points into pods via custom annotation dynamically.

    Bioinformatics pipelines have deep roots in HPC and assume local access to files on shared file systems.
    Containerised pipelines can now access cloud-native data stores via DLF from various sources via various protocols (for example GCS, S3 and HTTP archive).
    Both non-ML pipelines and newer ML pipelines can make use of Datasets created by DLF statically or dynamically, and have mount points injected via annotations.
    The use of Dataset simplifies the pipelines, improves security comparing with Onedata.
    The performance gets further enhanced with caching enabled.

    Dataset Lifecycle Framework addresses the requirement to access files locally by pipelines originated in HPC so that they can access cloud-native data stores on Kubeflow in Kubernetes environment directly.
    This integration is not only extremely useful in Bioinformatics but also potentially beneficial to all pipelines developed in HPC.
    It solves one of the major issues for HPC pipelines migrating into the clouds.

    \section{Acknowledgements}\label{sec:acknowledgements}

    The authors thank Dr. Tony Wildish for helpful discussions, as well as invaluable suggestions and edits on the manuscript.

    This project has received funding from the European Union's Horizon 2020 research and innovation programme "evolve" under grant agreement No 825061.
    It is also supported by the internal funding from European Bioinformatics Institute, European Molecular Biology Laboratory.

    
\end{document}